\documentclass[%
reprint,
superscriptaddress,
 amsmath,amssymb,
 aps,
prb,
longbibliography,
noeprint
]{revtex4-1}

\makeatletter
\renewcommand\frontmatter@abstractwidth{\dimexpr\textwidth-0.5in\relax}
\makeatother

\usepackage{graphicx}
\usepackage{dcolumn}
\usepackage{bm}
\usepackage{hyperref}

\usepackage{xcolor}
\hypersetup{
    colorlinks,
    linkcolor={red!50!black},
    citecolor={blue!50!black},
    urlcolor={blue!80!black}
}

\usepackage{helvet}

\newenvironment{myfont}{\fontfamily{phv}\selectfont}{\par}

\newcommand{\vect}[1]{\boldsymbol{#1}}
\newcommand{\Mvec}{\ensuremath{ \vect{M} }}
\newcommand{\mvec}{\ensuremath{ \vect{m} }}
\newcommand{\Hvec}{\ensuremath{ \vect{H} }}

\newcommand{\Hangle}{\ensuremath{ \phi_H }}
\newcommand{\Mangle}{\ensuremath{ \phi_M }}
\newcommand{\Mangleo}{\ensuremath{ \phi_M^{0} }}
\newcommand{\dMangleac}{\ensuremath{ \delta \phi_M^\mathrm{ac} }}

\newcommand{\JHM}{\ensuremath{ J_\mathrm{HM} }}
\newcommand{\JS}{\ensuremath{ J_\mathrm{S} }}
\newcommand{\Idc}{\ensuremath{ I_\mathrm{dc} }}
\newcommand{\RP}{\ensuremath{ R_\mathrm{P} }}
\newcommand{\RAP}{\ensuremath{ R_\mathrm{AP} }}
\newcommand{\DRGMR}{\ensuremath{ \Delta R_\mathrm{GMR} }}
\newcommand{\DRAMR}{\ensuremath{ \Delta R_\mathrm{AMR} }}
\newcommand{\Ro}{\ensuremath{ R_0 }}
\newcommand{\dRac}{\ensuremath{ \delta R_\mathrm{ac} }}

\newcommand{\Ic}{\ensuremath{ I_\mathrm{c} }}

\begin{document}

\begin{myfont}

\title{Giant magnetoresistance amplifier for spin-orbit torque nano-oscillators}

\author{\myfont Jen-Ru Chen}
\affiliation{\myfont Department of Physics and Astronomy, University of California, Irvine, California 92697, USA}
\author{\myfont Andrew Smith}
\affiliation{\myfont Department of Physics and Astronomy, University of California, Irvine, California 92697, USA}
\author{\myfont Eric A. Montoya}
\affiliation{\myfont Department of Physics and Astronomy, University of California, Irvine, California 92697, USA}
\author{\myfont Jia G. Lu}
\affiliation{\myfont Department of Physics and Astronomy and Department of Electrophysics, University of Southern California, Los Angeles, CA 90089, USA}
\author{\myfont Ilya N. Krivorotov}
\email{\myfont ilya.krivorotov@uci.edu}
\affiliation{\myfont Department of Physics and Astronomy, University of California, Irvine, California 92697, USA}


\begin{abstract}
\textbf{
Spin-orbit torque nano-oscillators based on bilayers of ferromagnetic (FM) and nonmagnetic (NM) metals are ultra-compact current-controlled microwave signal sources. They serve as a convenient testbed for studies of spin-orbit torque physics and are attractive for practical applications such as microwave assisted magnetic recording, neuromorphic computing, and chip-to-chip wireless communications. However, a major drawback of these devices is low output microwave power arising from the relatively small anisotropic magnetoresistance (AMR) of the FM layer. Here we experimentally show that the output power of a spin-orbit torque nano-oscillator can be enhanced by nearly three orders of magnitude without compromising its structural simplicity. Addition of a FM reference layer to the oscillator allows us to employ current-in-plane giant magnetoresistance (CIP GMR) to boost the output power of the device. This enhancement of the output power is a result of both large magnitude of GMR compared to that of AMR and different angular dependences of GMR and AMR. Our results pave the way for practical applications of spin-orbit torque nano-oscillators.
}
\end{abstract}


\maketitle

\end{myfont}

Electric current flowing in the plane of a  ferromagnetic (FM)/nonmagnetic (NM) bilayer can apply spin-orbit torque (SOT) to magnetization of the FM \cite{Ando2008, Miron2011, Liu2012, Hellman2017, Manchon2015, Freimuth2014, Belashchenko2019, Solyom2018}. The simplest example of such a SOT is spin Hall torque (SHT) arising from pure spin current in the NM layer that flows in the direction orthogonal to both the charge current and the FM/NM interface \cite{Sinova2015, Zhang2000, Hoffmann2013, Ou2019, Alghamdi2019}. When injected into the FM layer, this pure spin current applies SHT that can act as negative magnetic damping and thereby excite auto-oscillations of the FM magnetization \cite{Demidov2012, Liu2012a, Duan2014, Awad2017, Collet2016, Safranski2017, Tarequzzaman2019, Chen2016}.  The current-driven auto-oscillations of magnetization result in a microwave voltage generation by the FM/NM bilayer due to anisotropic magnetoresistance (AMR) of the FM\cite{Liu2012a, Liu2013}. Since AMR in thin films of FM metals is relatively small, the output microwave signal generated by the FM/NM bilayer spin Hall oscillators (SHOs) typically does not exceed several pW\cite{Liu2012a, Chen2016}. Here we report a new type of SHO with additional FM reference layer in which the microwave power generation relies on current-in-plane giant magnetoresistance (CIP GMR) \cite{Baibich1988, Binasch1989, Camley1989}. Since the magnitude of GMR significantly exceeds that of AMR, this new type of SHO generates significantly higher microwave power than the AMR-based SHOs. The maximum measured microwave power generated by the GMR SHO device exceeds 1~nW, which is nearly three orders of magnitude higher than the maximum microwave power produced by AMR SHO devices.

The main advantage of spin-orbit torque oscillators over spin transfer torque oscillators based on magnetic tunnel junctions (MTJs) \cite{Deac2008, Dussaux2010, Houssameddine2008, Zeng2013, Zeng2012, Skowronski2012}  is their structural simplicity and ease of fabrication. Indeed, a FM/NM bilayer spin-orbit torque oscillator device is powered by electric current flowing in the plane of the bilayer. Such a CIP nano-device can be produced by means of a single e-beam lithography step followed by a single etching step \cite{Evelt2018}. In contrast, MTJ-based oscillators are powered by electric current flowing perpendicular to the plane of the MTJ layers. Fabrication of such devices is a formidable task involving multiple lithography, etching, and deposition steps. Given the ease of fabrication of spin-orbit torque oscillators, they find use in fundamental studies of SOTs\cite{Safranski2019} as well as nonlinear magnetization dynamics \cite{Yang2015, Wagner2018} and hold promise for practical applications such as microwave-assisted magnetic recording \cite{Braganca2010}  and neuromorphic computing \cite{Torrejon2017}.  However, a major drawback of spin-orbit torque oscillators devices is low efficiency of converting direct bias current into microwave output signal. This poor conversion efficiency arises from the small value of AMR employed for converting current-driven magnetization auto-oscillations into electro-magnetic microwave signal. Here we present a realization of SHO that employs CIP GMR for generation of the microwave power, which significantly improves the conversion efficiency. 

\bigskip
\begin{large}
\noindent \textbf{\myfont Results}
\end{large}

\begin{figure*}[tbp]
\centering
\includegraphics[width=\linewidth]{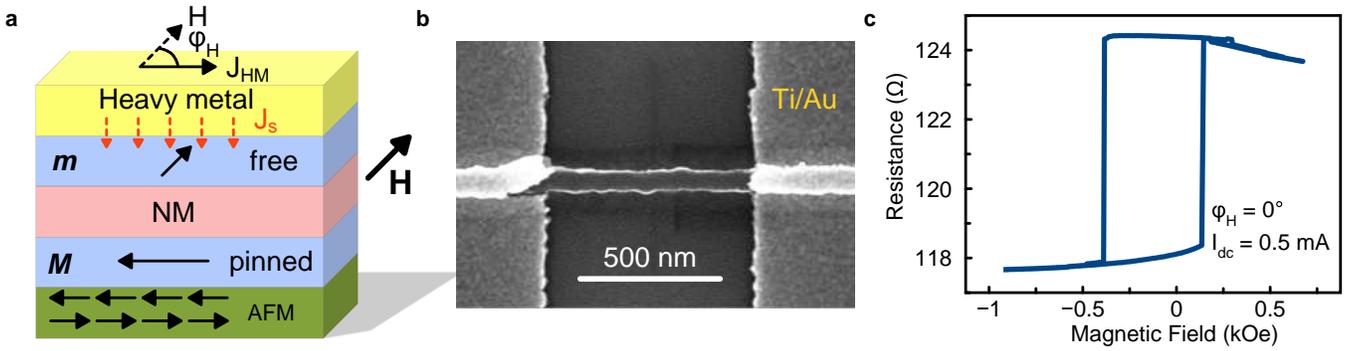}
\caption{\textbf{a} Schematic of a nanowire GMR SHO that consists of exchange biased CIP GMR spin valve, in which the free layer is interfaced with a heavy metal layer (Pt). Here $\Mvec$ is magnetization of the pinned layer, $\mvec$ is magnetization of the free layer, $\Hangle$ is the angle between electric current and direction of in-plane magnetic field $\Hvec$. Exchange bias field from the antiferromagnetic (AFM) layer is set in the direction of the electric current (along the nanowire axis). Direct electric current density $\JHM$ flowing in the Pt layer injects pure spin Hall current density $\JS$ into FM free layer. This spin current applies antidamping SOT to  and excites magnetization auto-oscillations. \textbf{b} Scanning electron micrograph (SEM) of the GMR SHO nanowire device. \textbf{c} Magnetoresistance of the nanowire GMR SHO device measured at T = 4.2 K with magnetic field applied parallel to the nanowire axis.}
\label{fig:1}
\end{figure*}

\noindent \textbf{\myfont Sample geometry and magnetoresistance.} Fig.~\ref{fig:1}a shows schematic of the GMR SHO device. The device is a nanowire made from AFM/FM/NM/FM/Pt exchange biased spin valve multilayer, where the direction of magnetization of the bottom FM layer is pinned by exchange bias field from the antiferromagnetic (AFM) layer \cite{Khanal2014, Ali2003, Krivorotov2003}. Direct electric current flowing along the nanowire in the heavy metal Pt layer applies SHT to magnetization of the adjacent free FM layer and excites its auto-oscillations \cite{Liu2012a}. CIP GMR in the FM/NM/FM spin valve serves as efficient converter of the FM magnetization auto-oscillations into resistance oscillations and microwave voltage resulting from these resistance oscillations \cite{Kiselev2003, Rippard2004, Krivorotov2005}.

The GMR SHO nanowire devices studied here were patterned from a (sapphire substrate)/Ir$_{25}$Mn$_{75}$(4~nm)/Co(2~nm)/Cu(4~nm)/ Co(0.5~nm)/Py(3.5~nm)/Pt(5~nm) multilayer deposited by magnetron sputtering. The 0.5 nm thick Co dusting layer was inserted between Cu and $\mathrm{Py} \equiv \mathrm{Ni}_{80}\mathrm{Fe}_{20}$ layers to enhance CIP GMR of the spin valve \cite{Parkin1993}. This metallic spin valve multi-layer was post-annealed at 523 K for 1 hour to set the direction of the exchange bias field parallel to the nanowire axis. A 65~nm wide by 40~$\mu$m long nanowire was patterned from the multilayer by using e-beam lithography and Ar ion milling. Two Ti (5~nm)/Au(40~nm) contact pads separated by a 740~nm wide gap were attached to the nanowire in order to apply in-plane electric bias current $\Idc$ to the wire. The 740~nm wide part of the nanowire between the contact pads forms the active region of the SHO where electric current density and resulting antidamping SHT can reach sufficiently high values to cancel the natural magnetic damping of the FM layer and induce magnetization auto-oscillations \cite{Liu2012a}. Fig. \ref{fig:1}b shows the scanning electron micrograph (SEM) of the GMR SHO device.

In order to compare performance of the GMR SHO to that of the conventional AMR SHO, we also fabricated and studied an AMR-based SHO with nominal lateral dimensions identical to those of the GMR SHO in Fig. \ref{fig:1}b. This reference AMR SHO was patterned from (sapphire substrate)/Cu(4 nm)/Co(0.5 nm)/Py(3.5 nm)/Pt(5 nm) magnetic multilayer. The 4 nm Cu underlayer is added to the standard AMR SHO design \cite{Liu2012a} to produce Oersted field acting on the free layer due to electric current in the Cu layer that is similar to that in the GMR SHO.

All measurements reported in this paper are performed in a continuous flow $^4$He cryostat at the bath temperature T = 4.2 K. Fig. \ref{fig:1}c shows resistance of the nanowire GMR SHO measured for magnetic field applied parallel to the wire axis at a small probe current $\Idc = 0.5$ mA. The data reveal switching of the free layer magnetization between parallel (low resistance, $ \RP = 117.8 \, \Omega$) and antiparallel (high resistance, $\RAP = 124.3 \, \Omega $) orientations with respect to magnetization of the pinned layer. The GMR ratio of the device $\DRGMR / \RP$, where $\DRGMR = \RAP - \RP$, is measured to be 0.055. We also measured AMR ratio in the reference AMR SHO device and found it to be 0.004.

\bigskip
\noindent \textbf{\myfont Microwave emission measurements.} We next perform measurements of microwave signal generation by the GMR SHO and AMR SHO driven by application of a sufficiently large $\Idc$ to the nanowire. For these measurements, we applied magnetic field H = 800 Oe in the plane of the sample at angle $\Hangle$ with respect to the nanowire axis as shown in Fig.~\ref{fig:1}a. This field is sufficiently high to rotate the free layer magnetization, while the pinned layer magnetization in the GMR SHO still has a large component parallel to the nanowire axis. For $\Hangle$ near $90^\circ$, the antidamping action of SHT is maximized because spin Hall current polarization is nearly opposite to the free layer magnetization \cite{Tarequzzaman2019}.

\begin{figure*}[htb]
\centering
\includegraphics[width=\linewidth]{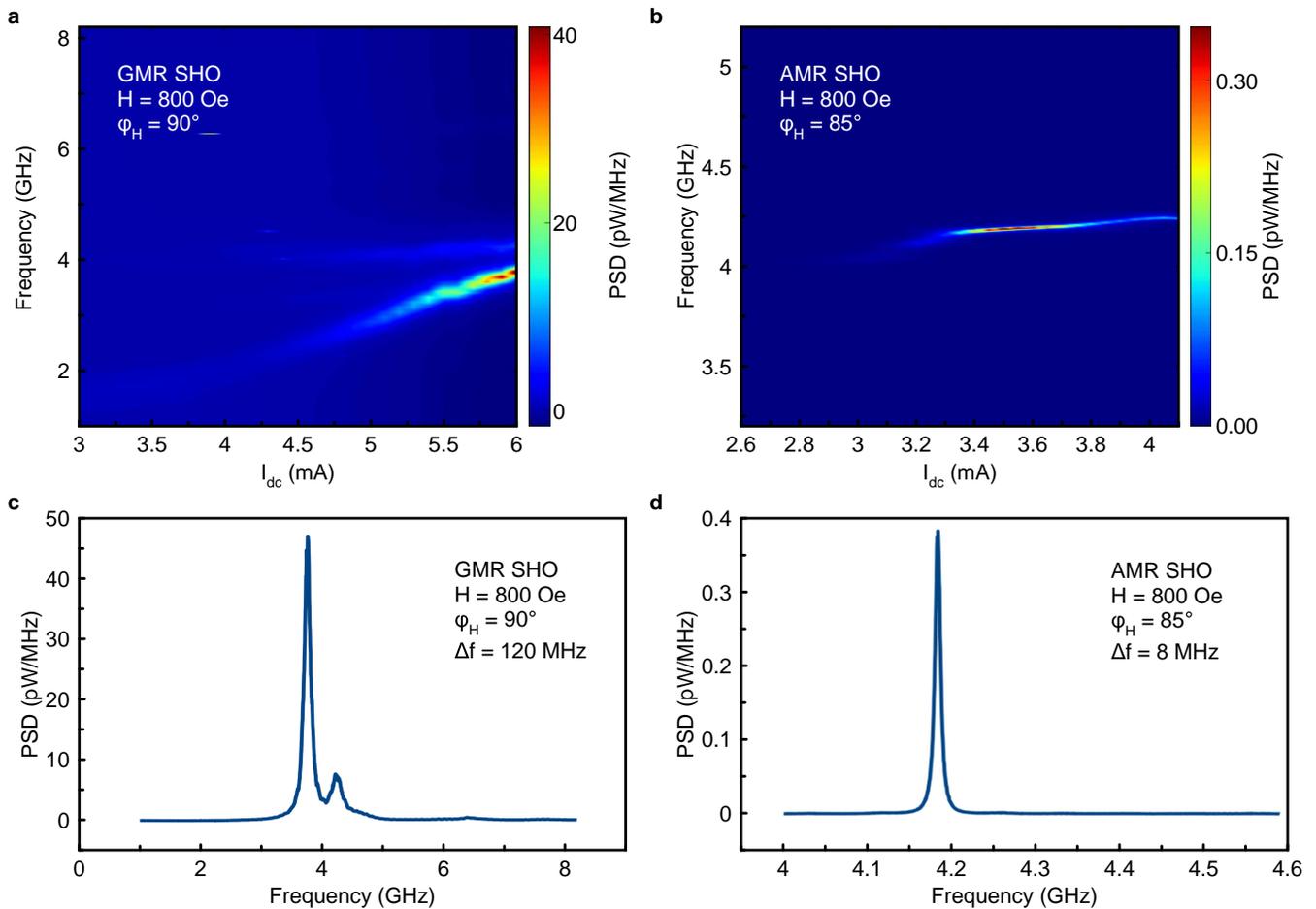}
\caption{ Spectral power density (PSD, color scale on the right) of microwave signal generated by GMR SHO \textbf{a} and AMR SHO \textbf{b}, measured as a function of direct current $\Idc$ applied to the nanowire. For these measurements, in-plane magnetic field H = 800 Oe was applied at angle $\Hangle$ with respect to the nanowire axis. PSD versus frequency for GMR SHO \textbf{c} and AMR SHO \textbf{d} measured at H = 800 Oe ($\Idc = 6.0$ mA and $\Hangle = 90^\circ$ for the GMR SHO, $\Idc = 3.65$ mA and $\Hangle = 85^\circ$ for the AMR SHO).}
\label{fig:2}
\end{figure*}

The data in Fig.~\ref{fig:2} shows measured power spectral density (PSD) of the microwave signal generated by the devices as a function of $\Idc$. These measurements were made using a microwave spectrum analyzer and a low noise microwave amplifier \cite{Liu2012a}. The data in Fig.~\ref{fig:2}b reveal that auto-oscillations of magnetization in the AMR SHO turn on for $\Idc$ exceeding the critical value of approximately 3.0 mA. For these measurements, we misalign the applied field direction from that perpendicular to the nanowire by $5^\circ(\Hangle = 85^\circ)$ in order to achieve significant conversion efficiency of auto-oscillations of magnetization into a microwave signal due to AMR \cite{Liu2012a, Chen2016}. The auto-oscillatory mode frequency exhibits a small shift to higher values with increasing bias current. This blue frequency shift is to be contrasted to the red frequency shift observed for bulk spin wave auto-oscillatory modes in Py/Pt SHOs \cite{Liu2012a}. We attribute this blue frequency shift in our Cu/Co/Py/Pt SHO to the Oersted field in the highly conductive Cu layer. This Oersted field points in nearly the same direction as the external applied field and thus increases the ferromagnetic resonance frequency\cite{Montoya2014} of the free layer with increasing $\Idc$.

Fig.~\ref{fig:2}a reveals that the GMR SHO device exhibits two auto-oscillatory modes above the critical current of approximately 4.0 mA. The higher frequency (HF) mode has nearly the same frequency as that of the AMR SHO and exhibits nearly the same blue frequency shift as the single auto-oscillatory mode of the AMR SHO. We thus identify this mode as the auto-oscillatory bulk spin wave mode of the free layer. As was shown in previous studies of AMR SHO, the amplitude of the auto-oscillatory bulk spin wave mode is maximized in the middle of the nanowire \cite{Liu2012a}. The other auto-oscillatory mode of GMR SHO appears at a lower frequency (LF) and exhibits a much larger blue frequency shift with increasing current compared to the HF mode. This strong blue frequency shift cannot be explained by the Oersted field from $\Idc$. Such a high blue shift has been previously observed for auto-oscillatory edge spin wave modes in AMR SHOs \cite{Liu2012a}. The edge spin wave mode results from spatially inhomogeneous demagnetizing field at the nanowire edges that produces a magnetic potential well for spin wave excitations \cite{Jorzick2002, Duan2015}. The edge mode in nanowires exhibits maximum amplitude at the wire edge. Recent theoretical work demonstrated that this blue frequency shift of the edge mode can result from non-linearity of the confinement potential leading to magnon repulsion \cite{Dvornik2018} . We therefore identify the LF mode as the edge mode of the free layer. We rule out the possibility of the LF mode being the pinned layer auto-oscillatory mode driven by spin Hall torque from the Ir$_{25}$Mn$_{75}$ layer \cite{Tshitoyan2015, Zhang2016, Saglam2018, Zhou2019} because strong exchange bias field acting on the pinned Co layer requires its auto-oscillatory frequency to be much higher than 2~GHz seen in Fig.~\ref{fig:2}a at the onset of auto-oscillations. 

\begin{figure*}[tbp]
\centering
\includegraphics[width=\linewidth]{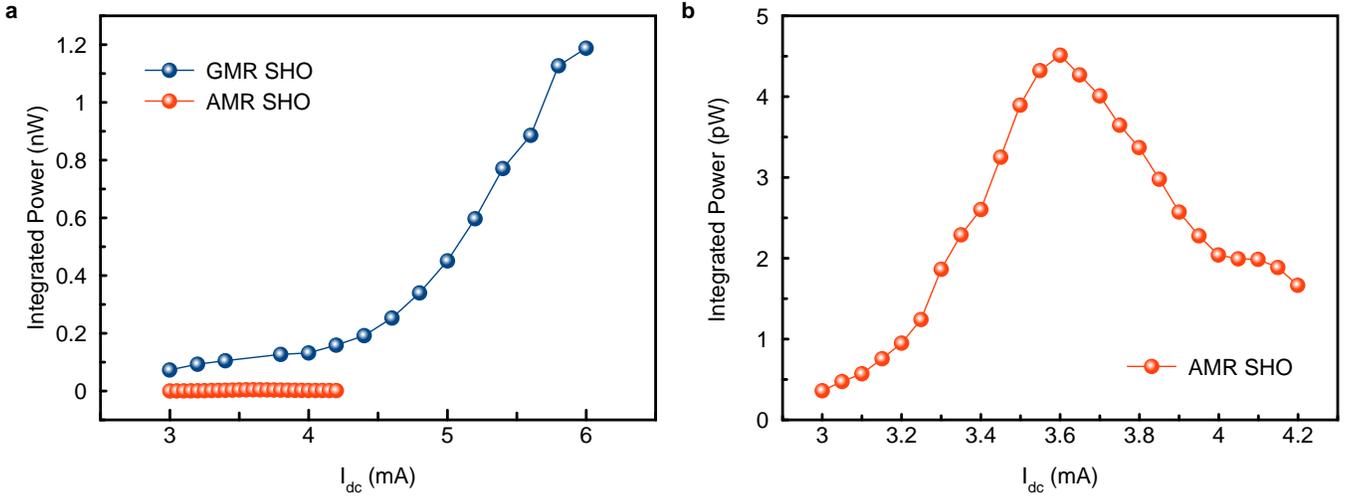}
\caption{ \textbf{a} Integral microwave power generated by GMR SHO (blue) and AMR SHO (red) as a function of bias current $\Idc$. \textbf{b} Rescaled axes of the AMR SHO data from \textbf{a}.}
\label{fig:3}
\end{figure*}

The microwave power emitted by the edge mode rapidly increases with increasing current bias and the overall microwave power emitted by the GMR SHO device is dominated by the edge mode. The dominant character of the edge mode auto-oscillations in the GMR SHO compared to the AMR SHO is likely a result of spatially inhomogeneous stray field from the pinned FM layer. Indeed, application of external magnetic field perpendicular to the nanowire axis $(\Hangle = 90^\circ)$ rotates the pinned layer magnetization towards the applied field direction. This gives rise to a stray field from the pinned layer that is opposite to the applied field near the edge of the free layer. This spatially inhomogeneous stray field enhances the localizing spin wave potential for the free layer edge mode. This, in turn, increases the spatial extent of the edge mode, which boosts the microwave power generated by the mode. Further theoretical studies are needed to test this proposed mechanism of the edge mode amplitude enhancement in GMR SHOs.

Fig.~\ref{fig:2}c and Fig.~\ref{fig:2}d show constant-current cuts of the data in Fig.~\ref{fig:2}a and Fig.~\ref{fig:2}b. These data reveal that the microwave power generated by the GMR SHO is much higher than that emitted by the AMR SHO. It is also clear that the spectral linewidth of the signal emitted by the AMR SHO (full width at half maximum $\Delta f$ = 8 MHz) is much smaller than that of both the HF and LF modes in the GMR SHO ($\Delta f$ = 120 MHz for the dominant LF mode). The higher spectral linewidth in the GMR SHO can be explained by three factors. First, since the critical current in the GMR SHO is higher, higher ohmic heating in GMR SHO leads to temperature-induced broadening of the spectral linewidth \cite{Slavin2009, Sankey2005, Boone2009}. Second, since two auto-oscillatory modes are simultaneously exited in the GMR SHO, interaction between these modes leads to spectral broadening of both modes \cite{Krivorotov2008, Iacocca2014}. Third, strong non-linear frequency shift of the edge mode gives rise to non-linear enhancement of the linewidth \cite{Zhou2019}.

\bigskip
\noindent \textbf{\myfont Bias dependence of SHO emission.}\begin{figure*}[ht]
\centering
\includegraphics[width=\linewidth]{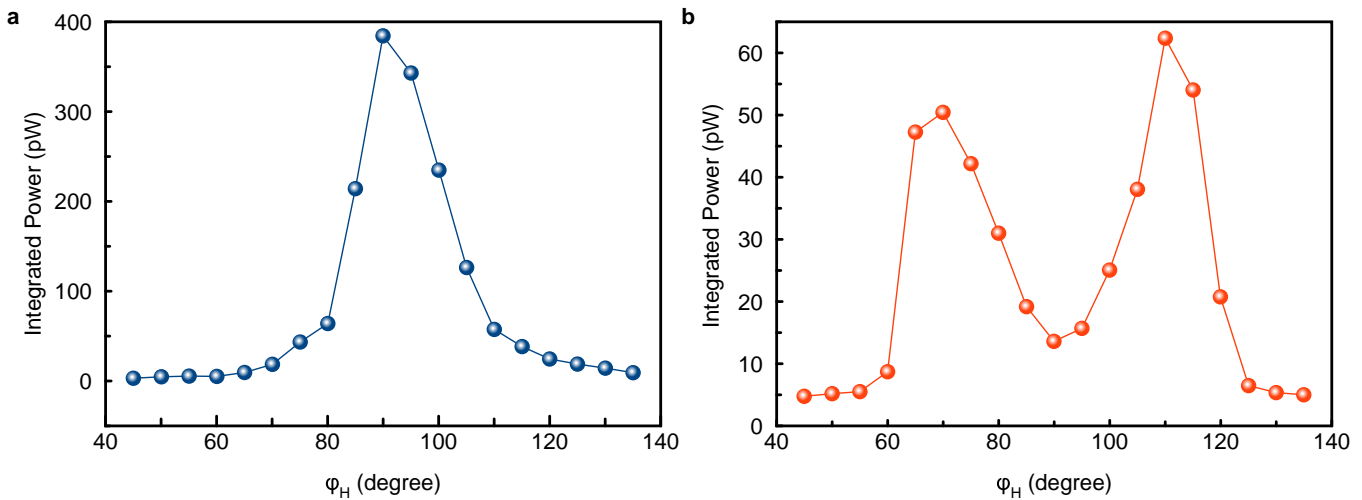}
\caption{ Dependence of the integrated microwave power generated by GMR SHO \textbf{a} and AMR SHO \textbf{b} on the direction $\Hangle$ of in-plane magnetic field H = 500 Oe. }
\label{fig:4}
\end{figure*}

Fig.~\ref{fig:3}a shows integrated microwave power emitted by the GMR SHO $(\Hangle = 90^\circ)$ and AMR SHO $(\Hangle = 85^\circ)$ devices measured at H = 800 Oe as a function of current bias $\Idc$. The data in Fig. 3 are obtained via integration of the spectra such as those in Fig.~\ref{fig:2}c and Fig.~\ref{fig:2}d. It is clear from Fig.~\ref{fig:3} that the output power of the GMR SHO is much higher than that of the AMR SHO for all bias current values. Fig.~\ref{fig:3}b shows the AMR SHO data from Fig.~\ref{fig:3}a with rescaled axes. Consistent with previous studies \cite{Liu2012a}, integrated power of the AMR SHO first increases and then decreases with increasing $\Idc$. The decrease of the integrated power at high currents can be attributed to enhanced magnon population and resulting strong non-linear magnon scattering at high current densities \cite{Demidov2011}. The integrated power of the GMR SHO monotonically increases with current up to the highest bias current value employed in this study ($\Idc$ = 6 mA). This is likely due to the higher critical current of the GMR SHO compared to the AMR SHO so that decrease of power induced by non-linear interactions is expected at $\Idc > 6$ mA.

\bigskip
\noindent \textbf{\myfont Anglular dependence of SHO emission.} We next study angular dependence of the microwave power generated by the GMR SHO and AMR SHO devices. Fig.~\ref{fig:4} shows the dependence of the integrated power emitted by GMR SHO (Fig.~\ref{fig:4}a) and AMR SHO (Fig.~\ref{fig:4}b) on the direction $\Hangle$ of a 500 Oe in-plane magnetic field. The maximum power generated by the GMR SHO is observed for magnetization direction $\Mangle$ perpendicular to the nanowire axis $(\Mangle = 90^\circ)$. This result can be explained by the angular dependence of CIP GMR \cite{Binasch1989} in this structure: $R = \RP + \DRGMR \cos(\Mangle)$. The output microwave power is proportional to square of the SHO current-driven resistance oscillation amplitude, $\dRac^2$, defined by $R(t) = \Ro + \dRac \sin(\omega t)$. The maximum output power is expected for the equilibrium direction of magnetization $\Mangleo$, which maximizes the amplitude of resistance oscillations $\dRac$. Substituting the expression for time dependence of the in-plane direction of magnetization $\Mangle (t) = \Mangleo + \dMangleac \sin( \omega t)$ into the expression for the angular dependence of CIP GMR and assuming $\dMangleac \ll 1\,(57^\circ)$, we derive $\dRac = -\DRGMR \sin(\Mangleo) \dMangleac$. It is thus clear that $\dRac^2$  and the output power of GMR SHO are maximized for $\Mangleo = \pi/2 \,(90^\circ)$, consistent with the experimental data in Fig.~\ref{fig:4}a. 

In the case of AMR SHO, the angular dependence of resistance is given by $ R = \RP - \DRAMR \cos^2 (\Mangle) $. Substituting the expression for $\Mangle(t)$ into the expression for the angular dependence of AMR and assuming $\dMangleac \ll 1$, we derive $\dRac = \DRAMR \sin(2 \Mangleo) \dMangleac$. Therefore, the maximum of  and the output microwave power of the AMR SHO can be expected at $\Mangleo = \pi/4 \,(45^\circ)$ and $\Mangleo = 3\pi/4 \,(135^\circ)$. The data in Fig.~\ref{fig:4}b reveal that the maximum power is observed near $\Mangleo = 70^\circ$ and $\Mangleo = 110^\circ$. This apparent discrepancy between theory and experiment can be explained by two considerations. First, the shape anisotropy field of the nanowire tends to pull magnetization of the free layer closer to the axis of the nanowire than the applied field direction $\Mangleo$. Therefore, $\Mangleo < 70^\circ$ for $\Hangle = 70^\circ$ ($\Mangleo > 110^\circ$ for $\Hangle = 110^\circ$). Second, the critical current $\Ic$ for excitation of auto-oscillations by SHT depends on the direction of magnetization \cite{Awad2017,Collet2016} $(\Ic \sim 1/\sin(\Mangleo))$ and $\Ic$ increases when magnetization rotates away from $\Mangleo = 90^\circ$. Therefore, rotation of the applied field away from $\Hangle = 90^\circ$ at a constant bias current $\Idc$ results in $\Idc < \Ic$ when $|\Hangle -90^\circ|$ reaches a certain critical value. This observation is consistent with the data in Fig.~\ref{fig:4}b that reveal a precipitous drop of the emitted power down to the background value for  $\Hangle < 60^\circ$ and $\Hangle > 120^\circ$.

\bigskip
\begin{large}
\noindent \textbf{\myfont Discussion}
\end{large}

In addition to SHT from the Pt layer, the free layer of the GMR SHO may experience spin transfer torque applied by spin current arising from the pinned Co layer \cite{Taniguchi2015, Gibbons2018, Baek2018, Amin2016}. However, we expect this torque to be relatively small in our nanowire system because magnetization of the pinned layer is nearly collinear with the electric current direction in the pinned Co layer. Anomalous spin-orbit torque  acting on the free layer is expected to be small for magnetization nearly perpendicular to the electric current direction and for the relatively small thickness of the free layer used in our studies, and thus can be neglected for our system as well.

In the case of GMR SHO, one particular direction of magnetization $(\Mangleo = \pi/2 \,(90^\circ))$ simultaneously maximizes the output microwave power and minimizes the critical current for the excitation of auto-oscillations of magnetization. Therefore GMR SHO offers good performance in terms of both the critical current and the output power.  In the case of AMR SHO, the critical current is minimized at $(\Mangleo = \pi/2 \,(90^\circ))$ while the amplitude of resistance oscillations is maximized at different angles $(\Mangleo = \pi/4 \,(45^\circ))$ and $(\Mangleo = 3\pi/4 \,(135^\circ)$. Therefore, AMR SHO design considerations necessarily include a trade-off between low critical current and high output power.  
	
In conclusion, we experimentally demonstrated that current-in-plane giant magnetoresistance can be used to enhance the output microwave power of a spin Hall oscillator by nearly two orders of magnitude compared to spin Hall oscillators that utilize anisotropic magnetoresistance for generation of microwave signal. Our data reveal that spin Hall oscillators with giant magnetoresistance signal amplification can be designed to simultaneously minimize the critical current for excitation of auto-oscillations and maximize the output microwave power of the oscillator. This enhancement increases the viability of spin Hall oscillators for emerging nanotechnology applications, such as neuromorphic and reservoir computing \cite{Torrejon2017, Romera2018, Markovic2019} and chip-to-chip wireless communications \cite{Lee2019}.

%

\bigskip
\begin{large}
\noindent \textbf{\myfont Acknowledgements}
\end{large}

\noindent This work was supported by the National Science Foundation through Grants No.\,DMR-1610146, No.\,EFMA-1641989 and No.\,ECCS-1708885. We also acknowledge support by the Army Research Office through Grant No.\,W911NF-16-1-0472, Defense Threat Reduction Agency through Grant No.\,HDTRA1-16-1-0025 and the Beall Innovation Award at the University of California, Irvine.

\bigskip
\begin{large}
\noindent \textbf{\myfont Author contributions}
\end{large}

\noindent 
J.R.C. made magneto-resistance and microwave emission measurements. J.R.C. and A.S. developed sample nanofabrication techniques and made the samples. J.R.C. and E.A.M. analyzed the data. I.N.K., J.G.L. and E.A.M. wrote the manuscript. I.N.K. and J.G.L. planned the study. I.N.K. managed the project. All authors discussed the results.

\end{document}